\documentclass[fleqn,10pt]{wlscirep}
\usepackage[utf8]{inputenc}
\usepackage[T1]{fontenc}
\usepackage{cite}

\newcommand{\COR}[1]{\textcolor{black}{ #1}}

\title{Characterisation of a single photon event camera for quantum imaging}

\author[1]{Victor Vidyapin}
\author[2,1]{Yingwen Zhang}
\author[1,*]{Duncan England}
\author[1,2]{Benjamin Sussman}
\affil[1]{National Research Council of Canada, 100 Sussex Drive, Ottawa, Ontario, K1A 0R6, Canada}
\affil[2]{Department of Physics, University of Ottawa, Ottawa, Ontario, K1N 6N5, Canada}

\affil[*]{duncan.england@nrc.ca}

\begin{abstract}
We show a simple yet effective method that can be used to characterize the per pixel quantum efficiency and temporal resolution of a single photon event camera for quantum imaging applications. Utilizing photon pairs generated through spontaneous parametric down-conversion, the detection efficiency of each pixel, and the temporal resolution of the system, are extracted through coincidence measurements. We use this method to evaluate the TPX3CAM, with appended image intensifier, and measure an average efficiency of $7.4\pm 2$\,\% and a temporal resolution of 7.3\,ns. Furthermore, this technique reveals important error mechanisms that can occur in post-processing. We expect that this technique, and elements therein, will be useful to characterise other quantum imaging systems. 
\end{abstract}

\begin{document}

\flushbottom
\maketitle
%
%
\thispagestyle{empty}


\section*{Introduction}

Photonics is emerging as an important platform for future quantum technologies. Single photons can transmit quantum information at the speed of light, making them the natural choice for quantum communication~\cite{Brassard1984,Ekert91}, and quantum-enhanced imaging or detection~\cite{Genovese2016}. All such experiments begin with a source of non-classical light, proceed with the manipulation and transmission of the light, and conclude with its detection. To fully understand experimental results and avoid errors, it is vitally important that each link in this chain is appropriately characterized. In this paper, we focus on the characterization of a single-photon sensitive time-tagging camera for use in quantum imaging experiments. Because an anticipated use-case for the camera is in quantum imaging, it is logical that a non-classical light source is used to evaluate its performance. In this way, we directly evaluate the camera in the framework in which it will be used. 

The camera under characterization is the TPX3CAM from Amsterdam Scientific Instruments which is a $256\times256$ silicon pixel array, where the arrival time and position of each event on the detector is tagged with nanosecond (1.6\,ns FWHM) and micrometer resolution (55\,$\mu$m pixel pitch) respectively. This camera is not, by itself, single-photon sensitive; instead a fast image intensifier is appended to the camera to improve the sensitivity.  \COR{The combination of a TPX3CAM and image intensifier results in a device which can tag the arrival time and position of every incident photon with nanosecond-scale accuracy, on over 65,000 pixels. This combination of temporal and spatial resolution is currently unprecedented in single photon sensitive cameras leading to a recent surge quantum optics experiments with this detector} ~\cite{Zhang2020,Zhang2021,Zhang2022,Zhang2022snapshot,Gao2022,Svihra2020,Nomerotski2020,Nomerotski2020spatial}. However the use of an image intensifier leads to complications which must be carefully considered. When one photon strikes the intensifier it is amplified into a burst of O$(10^5)$ photons, which will illuminate a small cluster of pixels. Furthermore, due to thresholding effects, the registered arrival time will have a small variance depending on the number of photons hitting a pixel, causing each pixel in the cluster to register a slightly different arrival time. Careful processing is thus required to assign a single position and time for every such cluster. Therefore, unlike conventional single photons cameras such as avalanche photodiode (APD) arrays, the process of detecting a single photon involves both hardware and software. Consequently, full charachterization of the camera system must also involve both hardware and software post-processing as is discussed in detail in this manuscript. 

We characterize the camera using pairs of photons generated by spontaneous parametric down-conversion (SDPC). The idea of using two-photon emission to calibrate the efficiency of photodetectors was first conceived over 40 years ago~\cite{Klyshko1980,Malygin1981} and has developed into a dependable method for characterising single photon detectors~\cite{Ware2004}. The concept is as follows: the two output modes of a SPDC source are split onto two detectors, $A$ and $B$ and the number of detection events $N_a$ and $N_b$ are measured.  The number of coincident detection events $N_{a,b}$, when both detectors fire simultaneously, are also recorded. Because SPDC photons are always generated in pairs then every detection should be coincident so, for perfect detectors: $N_{a,b} = N_a = N_b$. Deviations from this indicate imperfect detection efficiency. This principle can be used to calculate the efficiency $\eta$ of each detector:
\begin{equation}
    \eta_a = \frac{N_{a,b}}{N_b}, \hspace{1cm} \eta_b = \frac{N_{a,b}}{N_a}.
\label{eq:efficinecy_single}
\end{equation}
In a practical setting, $\eta$ will include collection losses as well as the detector efficiency, and dark counts, so these losses must be carefully calibrated to extract the true detector efficiency~\cite{Ware2004}. While this methodology was originally designed for single-pixel detectors, it can be extended to multi-pixel cameras~\cite{Qi2016}. \COR{A major advantage of this approach is that it is self-contained and does not require any previously calibrated sources or detectors.} Furthermore, because it is known that the two photons are emitted simultaneously (within a sub-picosecond coherence time) this method can also be used to determine the temporal resolution of a detector~\cite{Kwiat1994}. Here we combine these two ideas to show that photon pairs can be used to measure the temporal resolution and the position-dependent efficiency of the camera system. We also show that this method can also be used to highlight two error mechanisms that can arise in post-production: {\em double detection} where one photon leads to two detection events, and {\em dead zones} where two photons arriving at a similar time and position get counted as a single event.

\begin{figure}[ht]
\centering
\includegraphics[width=16cm]{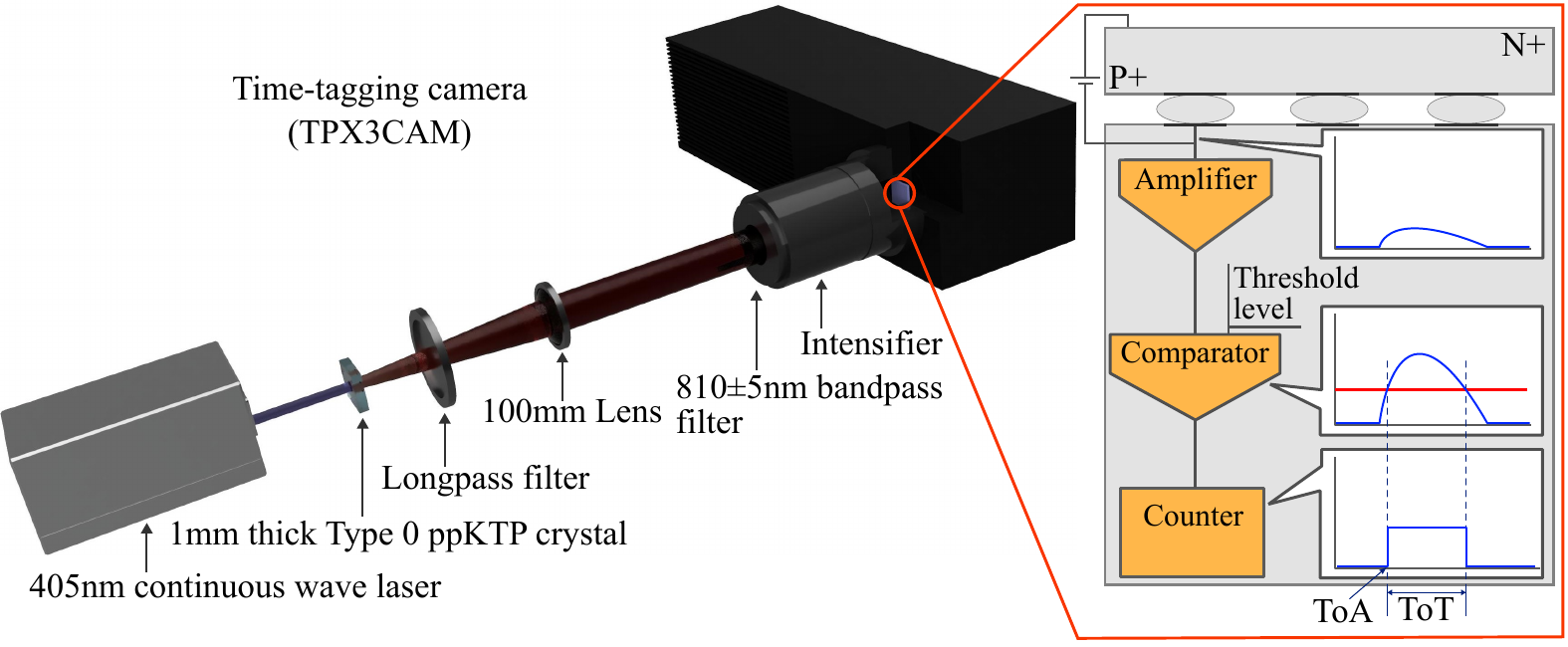}
\caption{A schematic diagram of the experiment. A 405\,nm continuous wave laser pumps a nonlinear crystal, generating pairs of photons by SPDC. A longpass filter removes the residual pump and the SPDC light is collimated by a lens. The photons impinge upon a image intensifier such that an incident photon is converted into a bright flash of light which can be registered on the time-tagging camera. A bandpass filter in front of the intensifier protects it from spurious light. {\bf Inset:} The charge on each pixel is amplified and compared to a threshold level to measure the time of arrival (ToA) and time over threshold (ToT) of each event. }
\label{fig:setup full}
\end{figure}

\section*{Results}

\subsection*{Apparatus}
A conceptual diagram of the experimental setup is shown in figure~\ref{fig:setup full}. A 405\,nm laser pumps a type-0 nonlinear crystal to produce degenerate pairs of photons centred at 810\,nm. The pump laser is blocked by a longpass interference filter. The photon pairs are collimated by a lens and sent to the camera system which consists of the TPX3CAM and appended intensifier, an additional bandpass filter ($810\pm5$\,nm) is attached directly to the front of the intensifier to remove stray light. Because the photons are imaged in the far field, the position that they are detected on the camera is proportional to the angle at which the photons are emitted. Due to conservation of momentum, degenerate photon pairs will ideally emerge from the crystal at equal but opposite angles so their arrival positions on the camera are anti-correlated such that $x_1\simeq -x_2$ and $y_1 \simeq -y_2$, where $x, y$ are the coordinates on the camera, and the center of the beam is at $x=y=0$. As discussed, the TPX3CAM is not single-photon sensitive, but it can be made so by attaching an image intensifier. When a photon is incident upon the intensifier, it first strikes the photocathode generating a photoelectron; the photoelectron is then amplified into a burst of electrons by a multichannel plate (MCP); the burst of electrons strikes a phosphor screen to generate a burst of photons, which are imaged onto the camera. The incident burst of photons is registered on a small cluster of camera pixels. \COR{Identifying clusters that are due to single photons, and attributing unique time and position coordinates to each cluster, requires careful post-processing of the data. }

Every pixel in the TPX3CAM contains: an amplifier which boosts the electrical signal; a comparator which compares this signal to a user-defined threshold; and a counter which logs the time at which the signal rises above the threshold, and the time at which falls below it. This process is shown pictorially in the inset of figure~\ref{fig:setup full}. \COR{The time that the amplified signal rises above the threshold is known as the time of arrival (ToA), and the time that the signal remains above threshold is known as the time over threshold (ToT). The ToA is a good approximation of when a pulse of light has struck the pixel, but it is not perfect. Because the ToA is registered only by the time the signal crosses the threshold, and not by a full digitization of the waveform, a brighter pulse of light will lead to a larger signal which will appear to have an earlier ToA, and a larger ToT. Conversely a weaker pulse of light will appear to have a later ToA and a shorter ToT~\cite{Nomerotski2019}. A cluster is brighter in the middle and weaker on the edges resulting in a spread in ToA values that can be $>100$\,ns~\cite{ianzano_fast_2020}.}

\subsection*{Post-processing}
\begin{figure} [ht]
\centering
\includegraphics[width=10cm]{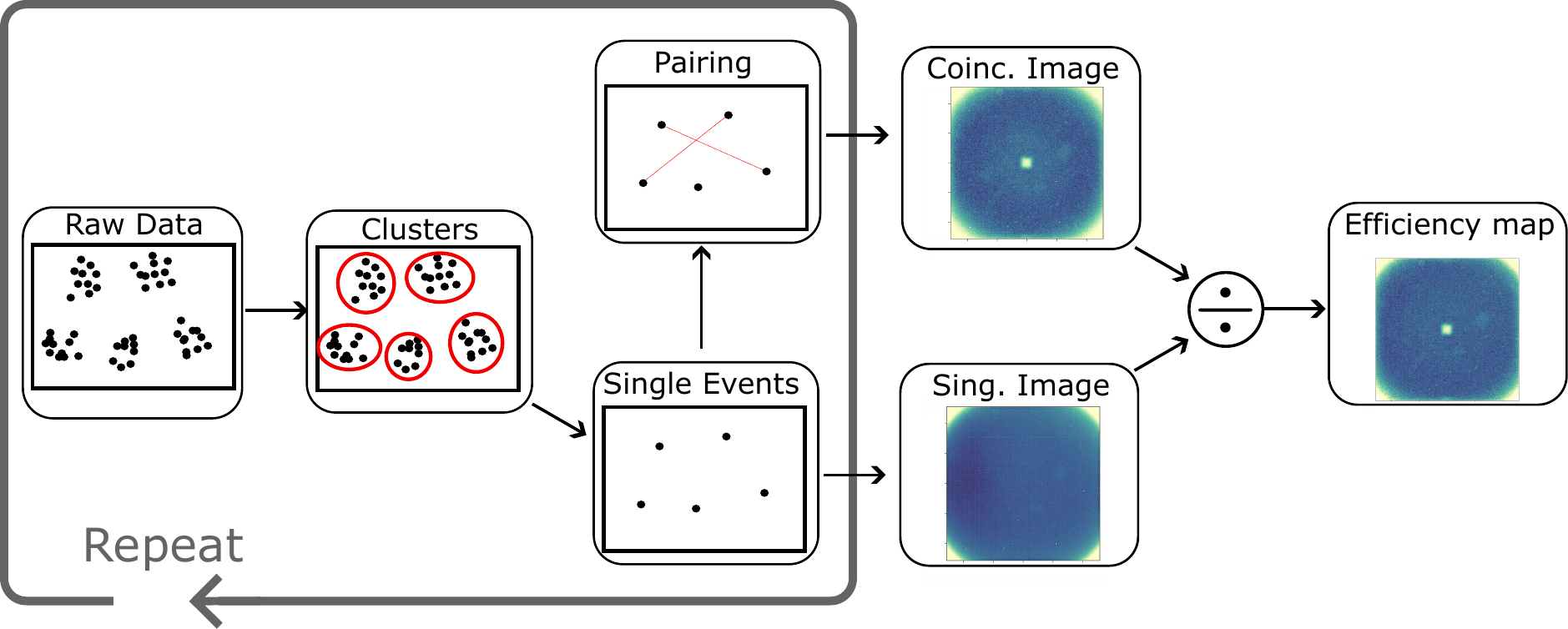}
\caption{\COR{Flow chart describing the processing steps required to measure the per-pixel efficieny. See text for details.}}
\label{fig:Flow_chart}
\end{figure}

 Converting the raw data stream into single events, and subsequently measuring the efficiency of the camera, requires a series of post-processing steps. These are described in detail in the below sections, and summarised by the flow-chart in figure~\ref{fig:Flow_chart}. 

\COR{The first step is {\em cluster identification} in which clusters of events are grouped together. Next, is {\em centroiding} where each cluster is assigned a single spatial-temporal coordinate. The sum of all these single events makes up the {\em singles image}. In the {\em pairing} step, pairs of events with correlated arrival time, and anti-correlated arrival positions are identified, an image containing the sum of all these pairs is known as the {\em coincidence image}. The {\em efficiency map} is a plot of the per-pixel efficiency of the camera which can be calculated by dividing the coincidence image by a rotated copy of the singles image.}

\subsection*{Cluster identification}
\COR{Because of the gain in the intensifier, a single incident photon is converted to a bright burst of photons. This burst illuminates multiple pixels resulting in a cluster of events which have similar, but not identical, spatial and temporal coordinates. The first step in the post-processing, which we term {\em cluster identification}, is the process of identifying clusters of events that are due to a single photon incident on the intensifier.} Typically the centre of the burst is brightest therefore the central pixel in the cluster will measure the highest intensity, and the surrounding pixels will be dimmer. Due to the nature of thresholding, the brightest central pixel will have the earliest ToA, and the surrounding pixels will have later ToAs. Therefore, even though the burst is due to a single photon, the pixels will appear to arrive at different times, which can vary by up to hundreds of nanoseconds. It is possible, to a degree, to correct for this temporal spread by so-called ToT-ToA correction~\cite{Frojdh2015}, but the correction is imperfect and significant temporal spread remains.

In order to identify clusters, we define a 3D box of size $\delta x \times \delta y \times \delta \tau$, where $x$, $y$, and $\tau$ are the positions and time of each event. A cluster is defined as a group of events that fall within a box of these dimensions. The exact dimensions of the box depend on the properties of the apparatus including the photon energy, intensifier gain, and camera threshold, and must be optimised for each system. In our case, the optimised values are $\delta x = \delta y = 17$\,pixels, and $\delta \tau = 300$\,ns. Although the $\delta \tau$ might seem excessive, there are noticeable improvements compared to 200\,ns or less.  This algorithm prioritizes efficiency over computing speed, and might not be as useful as other algorithms~\cite{Meduna2019} for experiments with a large number of events.

\begin{figure}[ht]
\centering
\includegraphics[width=12cm]{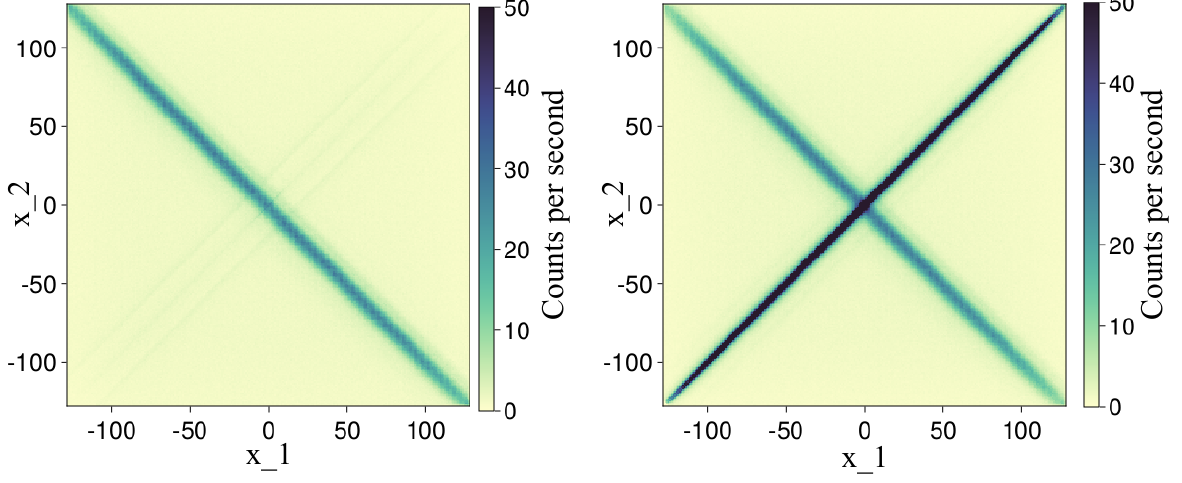}
\caption{Two-dimensional histograms of the horizontal position of pairs of events. Pairs of photons generated by SPDC exhibit strong anti-correlations resulting in an anti-diagonal line ($x_1 \simeq -x_2$). In {\bf (a)}, clustering has been performed appropriately and this anti-diagonal line dominates. In {\bf (b)}, the clustering algorithm parameters are chosen inappropriately and a strong diagonal line ($x_1\simeq x_2$) emerges because clustering can return two or more events for a single photon detection. }
\label{fig:doub_diag}
\end{figure}

If the clustering algorithm is performed incorrectly, for example if $\delta\tau$ is too small, then clusters can be split in half resulting in a single photon being counted twice, this has major implications for quantum imaging. This error mechanism, which we term double-counting, is illustrated graphically in figure~\ref{fig:doub_diag}. The camera is illuminated by the SPDC light and pairs of clusters are identified. A 2-dimensional histogram of the x position of one cluster $x_1$ is plotted against the x-position of the other cluster $x_2$. If the pair of clusters are due to genuine SPDC photons, then their position will be anti-correlated resulting in an anti-diagonal line on the 2-dimensional histogram. If the pair of clusters are due to double counting from improper clustering then their position will be correlated resulting in a diagonal line. \COR{In figure~\ref{fig:doub_diag}(a), the clustering algorithm has been appropriately applied with $\delta\tau = 300$\,ns. In this case the box is larger than the typical cluster size so all events from a cluster are included and the histogram is dominated by true coincidences from SPDC. In ~\ref{fig:doub_diag}(b), we see the effects of poor clustering. Here $\delta\tau$ is set to 17\,ns, which is significantly smaller than the typical cluster size, so clusters are split resulting in the possibility that two or more events can be generated by a single photon. In this case the histogram is dominated by false coincidences due to these split clusters.} Uncorrelated pairs arising in the off-diagonal areas occur due to various noise mechanisms including electrical noise and background light. Therefore we can immediately see, from a single plot, the effects of signal (anti-diagonal), noise (off-diagonal) and errors (diagonal) in our system.

\subsection*{Centroiding}
Once event clusters have been identified, we next need to assign a single position and arrival time to each cluster, this process is known as centroiding. Assigning the position is straightforward: we simply find the mean position of the cluster. In this case, we select the closest pixel to the mean, but it has been shown that centroiding can be used to improve spatial resolution by sub-dividing the pixels to gain a more accurate cluster center~\cite{Kim2020}. 

Assigning the arrival time is more complex. Because of the thresholding issue discussed above, each pixel in a cluster has a different ToA. In this case, computing the mean ToA is not a good choice because this can by greatly affected by low intensity pixels in the edge of the cluster. Three better options include assigning the ToA of the central pixel, the ToA of the pixel with the earliest ToA, or the ToA of the pixel which has the highest ToT. The result of these ToA correction methods are compared through a two photon coincidence measurement. Because the photons are produced in pairs, we know their arrival time should be identical to within their coherence time (sub-picosecond). Therefore, by measuring the arrival time difference across thousand of pairs, we can accurately measure the temporal resolution of the camera. This temporal resolution is affected by the ToA timing correction method.

\begin{figure}[ht]
\centering
\includegraphics[width=10cm]{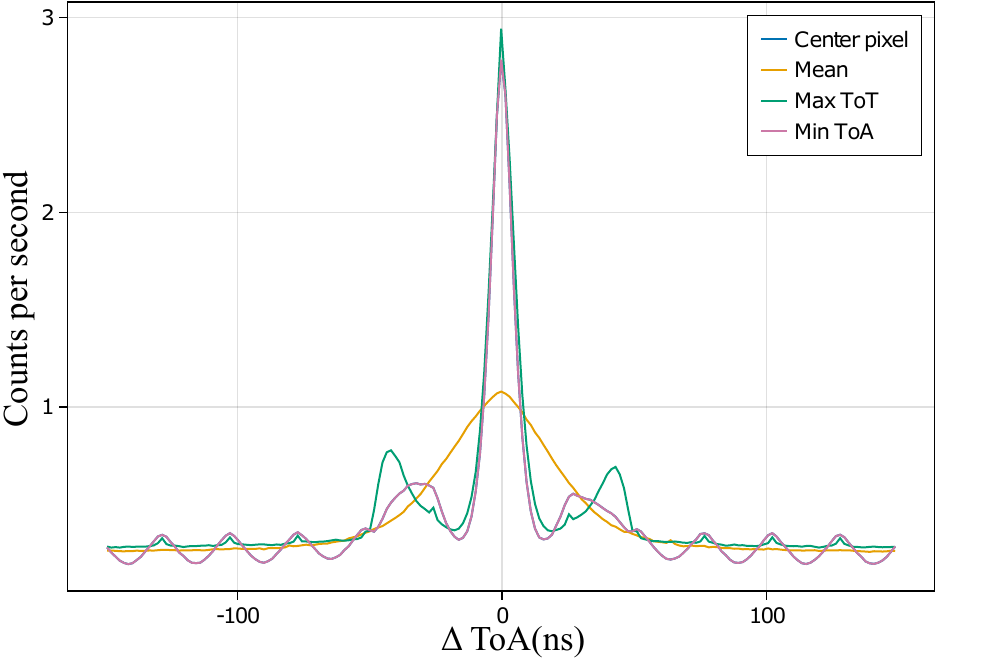}
\caption{Arrival time difference histogram of event pairs exhibiting a peak at $\Delta\text{ToA}=0$ due to correlated SPDC pairs. Since the coherence time of the photons is $\simeq100$\,fs, the width of this peak is a measure of the timing jitter of the system. Different methods have been used to assign the arrival time of a cluster, details of which are described in the text, resulting in different temporal performance. Note that the center pixel curve is not visible as it overlaps with the minimum ToA curve.}
\label{fig:split_sensor}
\end{figure}

To measure the temporal resolution of the camera, we split the detection events into two bins: one for photons that strike the top half of the sensor ($T$) and another for those that strike the bottom half ($B$). As the arrival positions of two photons in a pair are anti-correlated so, if one photon arrives in $B$, the other must arrive in $T$. Once the events have been binned, the arrival time difference between the photon pair is calculated as:
\begin{equation}
    \mathrm{\Delta ToA} =  \mathrm{ToA_T} -  \mathrm{ToA_B}.
\end{equation}
Figure~\ref{fig:split_sensor} shows a histogram of $\Delta$ToA values, where the ToA has been assigned to each cluster using one of the aforementioned methods. In each case, a peak centered at $\Delta$ToA$ = 0$ is indicative of the correlated nature of SPDC pair production. The width of the peak in the histogram returns the temporal resolution of the technique. As expected, assigning the mean ToA gives poor temporal resolution, but using the center, maximum ToT, and minimum ToA values all give comparable results. Assigning the ToA of the event with maximum ToT gives the highest resolution, so this method is used for the remainder of the results. The FWHM of the peak is 10.3\,ns, implying that the temporal resolution of the detector is $\frac{10.3}{\sqrt{2}} = 7.3$\,ns.

\subsection*{Efficiency Measurement}
As shown in equation~\ref{eq:efficinecy_single}, the efficiency of a pair of single photon detectors can be characterized by measuring coincident detection from an SPDC source. The same principles can be applied to characterize the efficiency of the camera. We begin by taking a 25\,second acquisition of the SPDC output resulting in $\sim 11$\,million detection events. This is used to generate a {\em singles image} $S(x,y)$ which is simply the total number of events detected on each pixel. Next we make another 25\,second acquisition which is identical to the first in every way except that the SPDC beam is blocked, this generates a {\em background singles image} $S_B(x,y)$. The {\em coincidence image} $C(x,y)$ is calculated by isolating pairs of events that are correlated in arrival time (to within the timing jitter measured in figure~\ref{fig:split_sensor}), and anti-correlated in position (to within the uncertainty shown in figure~\ref{fig:doub_diag}). Further details of coincidence selection technique can be found in the methods section. As well as true coincidences from SPDC pairs, the coincidence image also contains {\em accidental coincidences}. Sources of accidental coincidences include stray light, electrical noise, pairs in which the two photons derive from separate SPDC events, and combinations of the above processes. Considering both temporal and spatial coincidence windows, these false coincidences account for only $\sim 10^{-5}$ of the total number of coincidences in our setup, but are included in the efficiency measurement for completeness. In a different configuration, false coincidences may play a more significant role. Details of how the {\em background coincidence image} $C_B(x,y)$ is calculated can be found in the methods section.

\begin{figure}[ht]
\centering
\includegraphics[width=16cm]{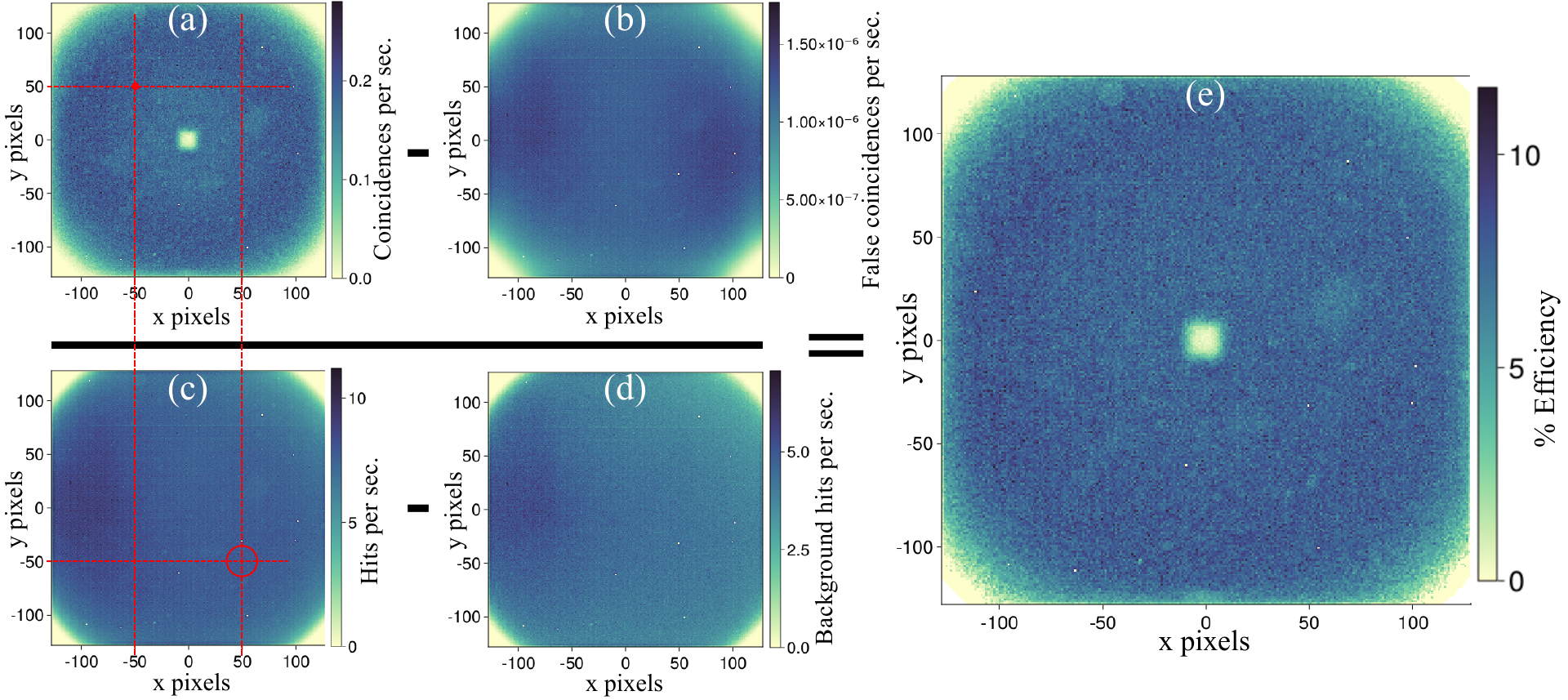}
\caption{Visualization of the calculation of the per-pixel efficiency of the sensor (equation~\ref{eq:efficiency}). {\bf (a)} A coincidence image is formed by including only events that have temporal and spatial partners. {\bf (b)} A background coincidences image calculated according to equation~\ref{eq:Cbg}. {\bf (c)} A raw image of the SPDC beam and background. {\bf (d)} Background image acquired with the SPDC beam switched off. {\bf (e)} The per-pixel efficiency is calculated by the ratio of the background-subtracted coincidences in a pixel (red dot in a) to the conjugate area in the singles background-subtracted image (red circle in c). }
\label{fig:efficiency}
\end{figure}

In direct analogy with equation~\ref{eq:efficinecy_single}, the camera efficiency can be calculated as the ratio of the background-subtracted coincidence image at a particular position, divided by the the background-subtracted singles image at the conjugate position. Because the position anti-correlations from the source are not perfect, the conjugate position is not a single pixel, but rather an area centered about the conjugate position whose diameter is determined by the width of the anti-diagonal line in figure~\ref{fig:doub_diag}. The position-dependent efficiency $\eta(x,y)$ is therefore given by:
\begin{equation}
    \eta(x,y) = \frac{C(x,y)-C_B(x,y)}{\overline{S}(-x,-y) - \overline{S_B}(-x,-y)},
    \label{eq:efficiency}
\end{equation}
where $\overline{S}(x,y)$ and $\overline{S_B}(x,y)$ are the singles and background singles images respectively, with a moving average applied to take into account the imperfect correlation. Details of this moving average can be found in the methods section. The resulting efficiency plot, shown in figure~\ref{fig:efficiency}(e), provides a map of the quantum efficiency of the entire detection system as a function of position. \COR{Note that a square area of $17\times 17$ pixels in the center of the camera appears to have zero efficiency. This dead zone is not a physical effect, but rather an effect of the post-processing: if two or more photons arrive at the intensifier at a similar position and time, then the clustering algorithm will combine both into a single event. Therefore, the choice of $\delta x$ and $\delta y$ in the clustering algorithm is important. If it is too small then clusters will be split and one photon will be registered as two events. If it is too large then two photons arriving close to each other will be registered as a single event. In this work, $\delta x = \delta y = 17$\,pixels achieved an appropriate compromise. }

The corners of the camera are not covered by the intensifier, so cannot be measured but the rest of the sensor, apart from the dead zone, is effectively characterised. In this region, the mean and standard deviation pixel efficiencies are $\left<\eta\right> = 7.4 \pm 2$\,\%. The efficiency is approximately uniform, though some less efficient patches are evident.

\section*{Discussion}

In this paper, we have described a method for characterising a single photon event camera using the time and momentum correlations inherent in photons pairs generated through SPDC. These methods were used to characterize a TPX3CAM event camera, which is made sensitive to single photons using an appended intensifier. In such a system, significant post-processing is required to analyse the data including cluster identification, centroiding and timing corrections. Accordingly, the performance of this system depends on both hardware and software, highlighting the importance of an effective characterization method.

We use this method to determine that the position-dependent quantum efficiency of the system is approximately uniform with a mean efficiency of $\left<\eta\right> = 7.4 \pm 2$\,\%, and the temporal resolution is 7.3\,ns FWHM. It is important to note that the reported values depend on the specific hardware and software settings used in our laboratory. As such, this paper should not be considered a universal characterisation of the TPX3CAM platform, but rather a recipe for evaluating such camera systems. \COR{The efficiency measurement includes losses from other optical elements: two filters and one lenses. In practice, at least one lens would be required to form an imaging system, and filters are also required to ensure the sensor is not overwhelmed by stray light, as such $\left<\eta\right>$ should be considered as the efficiency of the entire imaging system. The mean efficiency of the detector alone is calculated to be $\left<\eta_c\right> = 8.0\pm2$\% by dividing by the known transmission of these three elements (see methods section). }

We also identify an error mechanism that can occur in post-processing due to improper cluster identification, where a single photon is registered as two or more separate events. We also encounter the reverse problem, where multiple photons arriving at a similar time and position be counted as a single event. These are examples of how the proposed characterization method can evaluate both software as well as hardware performances.

\COR{While this methodology was developed in the framework of the TPX3CAM, we expect that elements therein will be applicable to other devices. For example, intensified CCD cameras will also experience clustering issues, and densely-packed single photon avalanche detector SPAD arrays~\cite{Sajeed2021} may experience cross-talk between adjacent pixels. The analysis shown in figure~\ref{fig:doub_diag} will be useful for diagnosing these issues. Single photon array detectors such as these,} are emerging as powerful tools for quantum enhanced imaging. The use of quantum-correlated light to characterize these cameras for quantum imaging is logical because it allows us to benchmark their performance directly in the setting in which they will be used.

\section*{Methods}

\subsection*{Setup}
A 405\,nm continuous-wave laser pumps a 1\,mm thick type-0 periodically poled potassium titanyl phosphate (ppKTP) crystal to produce \COR{around 2\,Million entangled photons pairs per second through SPDC, centred at 810\,nm}. The pump laser is then removed by a longpass interference filter \COR{(Thorlabs FELH0600)}. The photon pairs are collimated by a 10\,cm focal length lens \COR{(Thorlabs LA1509-B)} to illuminate the full sensor area of the camera system which consists of the TPX3CAM and appended image intensifier (Photonis Cricket with Hi-QE Red photocathode). An additional $810\pm5$\,nm bandpass filter \COR{(Thorlabs FELH0600)} is attached directly to the front of the intensifier to remove stray light and isolate degenerate SPDC photons. \COR{The transmission of the longpass filter, lens, and bandpass filter at 810\,nm are 96\%, 99.5\% and 97\% respectively, totalling 92.7\% overall transmission.}

\COR{While type-0 ppKTP was used for this demonstration, it is worth noting that other nonlinear crystals (for example, $\beta$-Barium Borate (BBO), potassium titanyl phosphate (KTP) lithium niobate (LiNbO$_3$) etc.)  would be equally appropriate.  The most important condition is that the crystal should produce photon pairs that are emitted in a non-colinear fashion to ensure that the two photons impinge upon different parts of the camera. In most SPDC sources this condition is automatically fulfilled due to phase matching conditions. In type-II crystals, the two photons will be generated with opposite polarization, which will introduce a time delay due to birefringence. However, this delay will be in the picosecond regime and will have little practical effect on the temporal characterisation of the camera.}

\subsection*{Cluster identification} 

\COR{Due to the gain of the intensifier, each photon detected results in a cluster of events on the camera. This cluster must be reduced to a single spatio-temporal event. For this purpose, a custom cluster identification algorithm that accounts for the different spatial and temporal ranges of a cluster was created.} The algorithm involves picking every element that is not already in a cluster, and comparing to the following 200 events, if an event is within a 17\,pixel $\times 17$\,pixel $\times 300$\,ns box then it belongs in the cluster and is not involved in further checks. \COR{Each cluster contains O(10) events which are distributed over $>100$\,ns due to large ToA delays from events at the edge of a cluster. Because the events from the camera are sorted temporally, it is possible that several clusters from different spatial positions will temporally overlap. By comparing each element to the following 200, we can be reasonably confident that every event in a cluster is identified. Comparing more than 200 elements will produce better results, but the small improvement does not justify the increased processing time. The spatial and temporal dimensions of the box (17\,pixles and 300\,ns respectively) were optimised by observing the efficiency and noise after centroiding; increasing the dimensions improves efficiency as more events are captured, but also more false events are captured so noise increases.}

There are many existing cluster identification algorithms that can run faster with satisfactory efficiency like DBSCAN or the algorithm developed by Meduna et al~\cite{Meduna2019}. Due to the non-linear relationship of $\text{ToT}$ to $\Delta \text{ToA}$ and the increased uncertainty in $\Delta\text{ToA}$ at smaller $\text{ToT}$, density based algorithms struggle with variable density within a cluster occasionally either count two clusters as one or one cluster as two resulting in errors~\cite{Frojdh2015}. For practical application such errors might be insignificant when compared to run-time gains. However, while density based algorithms might be suitable for running an experiment, the algorithm provided in this paper is more suitable for characterising the camera more accurately and as a benchmark for other algorithms. Furthermore, the freedom to tune the parameters and utilize the provided metrics gives insight into the shape and size of the clusters and properties of the intensifier. 

\subsection*{Efficiency measurement}

The efficiency measurement requires four constituent images, as shown in equation~\ref{fig:efficiency}. This section details the calculation of these four images.  

The coincidence image $C(x,y)$ is calculated in two steps: first pairs of events that are temporally correlated are identified, secondly these events are sifted to find pairs that are also spatially anti-correlated. Neither the temporal or the spatial (anti)correlations are perfect, so we define coincidence windows. From the width of the main peak in figure~\ref{fig:split_sensor}, the temporal coincidence window is chosen to be $\Delta\mathrm{ToA} = 20$\,ns. Ideally, the photons would emerge from the crystal with exactly opposite angles, however, due to uncertainties in the momentum of the pump photons and the slight wavelength difference between SPDC photons, an uncertainty in the momentum anti-correlation is observed as a width of $\sigma_x = 20$\,pixels for the anti-diagonal line seen in figure~\ref{fig:doub_diag} for the $x$-coordinate, and similarly in the $y$-coordinate. Therefore, the three conditions required to define a pair of coincidences are:
\begin{eqnarray}
\Delta\mathrm{ToA} &=& |\mathrm{ToA}_1 - \mathrm{ToA}_2| \leq 20\mathrm{\,ns},\\
\sigma_x &=& |x_1 + x_2| \leq 20\mathrm{\,pixels},\\
\sigma_y &=& |y_1 + y_2| \leq 20\mathrm{\,pixels}.
\end{eqnarray}
Here the $x$ and $y$ pixel labels have been shifted such that the center of the SPDC beam is at $x=y=0$. All events that do not have a spatial-temporal correlated partner are rejected, and the remaining events at each pixel are summed to form the 2-dimensional coincidence image $C(x,y)$.

We must also calculate the smoothed singles image $\overline{S}(x,y)$ and background singles image $\overline{S_B}(x,y)$. Because the momentum correlation between pairs of photons is not perfect, we cannot calculate the efficiency directly by comparing coincidences and singles from geometrically opposite pixels. Instead, the coincidences from a given pixel are divided by the mean of the pixels in an appropriate region in the singles image. As before, the radius of this region is 20\,pixels.  The smoothed singles image $\overline{S}(x,y)$ is therefore simply calculated by assigning each pixel the mean of the relevant area in the singles image. The same process is applied to the background image to form the smoothed background singles image $\overline{S_B}(x,y)$.

Finally, background coincidences due to background light, electrical noise, double-pair emission, and combinations of the above events, need to be eliminated. These accidental coincidences can be calculated from the singles images by assuming they arise from uncorrelated processes. In this case, the number of accidental coincidences between a pixel and the conjugate area can be calculated to return the background coincidence image $C_B(x,y)$:
\begin{equation}
    C_B(x,y) = S(x,y)\overline{S}(-x,-y)\times \Delta\text{ToA}.
    \label{eq:Cbg}
\end{equation}
Where $S(x,y)$ and $\overline{S}(-x,-y)$ are expressed in counts per second, and $\Delta$ToA$=20$\,ns is the coincidence window chosen to measure the coincidence image. In this setup, the background coincidence rate is $\sim10^{-5}$ times lower than the true coincidence rate so makes no practical difference to the efficiency measurement, but is included for completeness. 

\subsection*{Comparison}
\COR{In order to judge the validity of our measurements we describe two alternative techniques to measure the efficiency and temporal resolution.}
\subsubsection*{Efficiency}
\COR{To provide a comparative measure of the camera efficiency, we couple one mode of the SPDC source into a single mode fiber, and use a single pixel avalanche photodiode (APD) to measure the photon flux at the output of the fiber. In this case we measure the count rate to be $R_{{\mathrm ref}} = 83,500\pm300$\,counts/s. We then remove the fiber and illuminate a section of the intensifier (using the same filters as in figure~\ref{fig:setup full}) with the same flux and measure a count rate of $R_{\mathrm{cam}}=8,550\pm60$. The manufacturer-specified quantum efficiency of the APD is $\eta_{\mathrm ref} = 68\pm6$, we can therefore estimate the mean pixel efficiency of the camera to be $\left<\eta\right> = \eta_{\mathrm ref} \times R_{\mathrm{cam}}/R_{{\mathrm ref}}  = 7.0\pm0.7$\,\%. This value is in good agreement with the value that was more rigorously measured in the paper. Note that this type of measurement relies upon the manufacturer-quoted APD efficiency and is not a stand-alone measurement. }

\subsubsection*{Temporal resolution}
\COR{To provide a comparative measure of the temporal resolution, we use a pulsed laser whose pulse duration (150\,fs) is significantly shorter than the expected temporal resolution. We partition the laser into two beam and illuminate two spots on the camera. We then construct a timing difference histogram between the two regions. The result is a series of peaks, separated by the laser repetition period (12.5\,ns). The average and standard deviation of the FWHM of these peaks is $11.8\pm0.8$\,ns, implying that the temporal resolution is: $\frac{11.8\pm0.8}{\sqrt{2}} = 8.3\pm0.6$\,ns. Due to the fitting of multiple peaks, we expect this answer to be more accurate than the measurement in figure~\ref{fig:split_sensor}, however both measurements are in reasonable agreement suggesting that the photon pair technique is valid. }

\section*{Acknowledgements}

The authors \COR{would like to thank Guillaume Thekkadath for performing the temporal comparison measurement. They are also} grateful to Andrei Nomerotski, Hazel Hodgson, Denis Guay, and Doug Moffatt for stimulating discussions and technical support. This work was partly supported by Defence Research and Development Canada and the National Research Council's Quantum Sensors Challenge Program.

\section*{Author contributions statement}
D.E and Y.Z. conceived the experiment,  Y.Z. and V.V. conducted the experiment, V.V. and Y.Z. analysed the results. V.V., Y.Z., D.E. and B.S. wrote and reviewed the manuscript. 

\section*{Competing interests}

The authors declare no competing interests. 

\section*{Data availability }

The datasets used during the current study are available from the corresponding author on reasonable request.

\bibliography{CameraCal}

\begin{thebibliography}{10}
\urlstyle{rm}
\expandafter\ifx\csname url\endcsname\relax
  \def\url#1{\texttt{#1}}\fi
\expandafter\ifx\csname urlprefix\endcsname\relax\def\urlprefix{URL }\fi
\expandafter\ifx\csname doiprefix\endcsname\relax\def\doiprefix{DOI: }\fi
\providecommand{\bibinfo}[2]{#2}
\providecommand{\eprint}[2][]{\url{#2}}

\bibitem{Brassard1984}
\bibinfo{author}{Brassard, C.} \& \bibinfo{author}{Bennett, C.~H.}
\newblock \bibinfo{title}{Quantum cryptography: Public key distribution and
  coin tossing}.
\newblock In \emph{\bibinfo{booktitle}{International conference on computers,
  systems and signal processing}} (\bibinfo{year}{1984}).

\bibitem{Ekert91}
\bibinfo{author}{Ekert, A.~K.}
\newblock \bibinfo{journal}{\bibinfo{title}{Quantum cryptography based on
  bell's theorem}}.
\newblock {\emph{\JournalTitle{Phys. Rev. Lett.}}}
  \textbf{\bibinfo{volume}{67}}, \bibinfo{pages}{661--663}
  (\bibinfo{year}{1991}).

\bibitem{Genovese2016}
\bibinfo{author}{Genovese, M.}
\newblock \bibinfo{journal}{\bibinfo{title}{Real applications of quantum
  imaging}}.
\newblock {\emph{\JournalTitle{Journal of Optics}}}
  \textbf{\bibinfo{volume}{18}}, \bibinfo{pages}{073002}
  (\bibinfo{year}{2016}).

\bibitem{Zhang2020}
\bibinfo{author}{Zhang, Y.} \emph{et~al.}
\newblock \bibinfo{journal}{\bibinfo{title}{Multidimensional quantum-enhanced
  target detection via spectrotemporal-correlation measurements}}.
\newblock {\emph{\JournalTitle{Physical Review A}}}
  \textbf{\bibinfo{volume}{101}}, \bibinfo{pages}{053808}
  (\bibinfo{year}{2020}).

\bibitem{Zhang2021}
\bibinfo{author}{Zhang, Y.}, \bibinfo{author}{England, D.},
  \bibinfo{author}{Nomerotski, A.} \& \bibinfo{author}{Sussman, B.}
\newblock \bibinfo{journal}{\bibinfo{title}{High speed imaging of
  spectral-temporal correlations in hong-ou-mandel interference}}.
\newblock {\emph{\JournalTitle{Optics Express}}} \textbf{\bibinfo{volume}{29}},
  \bibinfo{pages}{28217--28227} (\bibinfo{year}{2021}).

\bibitem{Zhang2022}
\bibinfo{author}{Zhang, Y.}, \bibinfo{author}{Orth, A.},
  \bibinfo{author}{England, D.} \& \bibinfo{author}{Sussman, B.}
\newblock \bibinfo{journal}{\bibinfo{title}{Ray tracing with quantum correlated
  photons to image a three-dimensional scene}}.
\newblock {\emph{\JournalTitle{Physical Review A}}}
  \textbf{\bibinfo{volume}{105}}, \bibinfo{pages}{L011701}
  (\bibinfo{year}{2022}).

\bibitem{Zhang2022snapshot}
\bibinfo{author}{Zhang, Y.}, \bibinfo{author}{England, D.} \&
  \bibinfo{author}{Sussman, B.}
\newblock \bibinfo{journal}{\bibinfo{title}{Snapshot hyperspectral imaging with
  quantum correlated photons}}.
\newblock {\emph{\JournalTitle{arXiv preprint arXiv:2204.05984}}}
  (\bibinfo{year}{2022}).

\bibitem{Gao2022}
\bibinfo{author}{Gao, X.}, \bibinfo{author}{Zhang, Y.},
  \bibinfo{author}{D’Errico, A.}, \bibinfo{author}{Heshami, K.} \&
  \bibinfo{author}{Karimi, E.}
\newblock \bibinfo{journal}{\bibinfo{title}{High-speed imaging of
  spatiotemporal correlations in hong-ou-mandel interference}}.
\newblock {\emph{\JournalTitle{Optics Express}}} \textbf{\bibinfo{volume}{30}},
  \bibinfo{pages}{19456--19464} (\bibinfo{year}{2022}).

\bibitem{Svihra2020}
\bibinfo{author}{Svihra, P.} \emph{et~al.}
\newblock \bibinfo{journal}{\bibinfo{title}{Multivariate discrimination in
  quantum target detection}}.
\newblock {\emph{\JournalTitle{Applied Physics Letters}}}
  \textbf{\bibinfo{volume}{117}}, \bibinfo{pages}{044001}
  (\bibinfo{year}{2020}).

\bibitem{Nomerotski2020}
\bibinfo{author}{Nomerotski, A.}, \bibinfo{author}{Keach, M.},
  \bibinfo{author}{Stankus, P.}, \bibinfo{author}{Svihra, P.} \&
  \bibinfo{author}{Vintskevich, S.}
\newblock \bibinfo{journal}{\bibinfo{title}{Counting of hong-ou-mandel bunched
  optical photons using a fast pixel camera}}.
\newblock {\emph{\JournalTitle{Sensors}}} \textbf{\bibinfo{volume}{20}},
  \bibinfo{pages}{3475} (\bibinfo{year}{2020}).

\bibitem{Nomerotski2020spatial}
\bibinfo{author}{Nomerotski, A.} \emph{et~al.}
\newblock \bibinfo{journal}{\bibinfo{title}{Spatial and temporal
  characterization of polarization entanglement}}.
\newblock {\emph{\JournalTitle{International Journal of Quantum Information}}}
  \textbf{\bibinfo{volume}{18}}, \bibinfo{pages}{1941027}
  (\bibinfo{year}{2020}).

\bibitem{Klyshko1980}
\bibinfo{author}{Klyshko, D.}
\newblock \bibinfo{journal}{\bibinfo{title}{Use of two-photon light for
  absolute calibration of photoelectric detectors}}.
\newblock {\emph{\JournalTitle{Soviet Journal of Quantum Electronics}}}
  \textbf{\bibinfo{volume}{10}}, \bibinfo{pages}{1112} (\bibinfo{year}{1980}).

\bibitem{Malygin1981}
\bibinfo{author}{Malygin, A.}, \bibinfo{author}{Penin, A.} \&
  \bibinfo{author}{Sergienko, A.}
\newblock \bibinfo{journal}{\bibinfo{title}{Absolute calibration of the
  sensitivity of photodetectors using a biphotonic field}}.
\newblock {\emph{\JournalTitle{JETP Lett.}}} \textbf{\bibinfo{volume}{33}},
  \bibinfo{pages}{477--480} (\bibinfo{year}{1981}).

\bibitem{Ware2004}
\bibinfo{author}{Ware, M.} \& \bibinfo{author}{Migdall, A.}
\newblock \bibinfo{journal}{\bibinfo{title}{Single-photon detector
  characterization using correlated photons: the march from feasibility to
  metrology}}.
\newblock {\emph{\JournalTitle{journal of modern optics}}}
  \textbf{\bibinfo{volume}{51}}, \bibinfo{pages}{1549--1557}
  (\bibinfo{year}{2004}).

\bibitem{Qi2016}
\bibinfo{author}{Qi, L.}, \bibinfo{author}{Just, F.}, \bibinfo{author}{Leuchs,
  G.} \& \bibinfo{author}{Chekhova, M.~V.}
\newblock \bibinfo{journal}{\bibinfo{title}{Autonomous absolute calibration of
  an iccd camera in single-photon detection regime}}.
\newblock {\emph{\JournalTitle{Optics Express}}} \textbf{\bibinfo{volume}{24}},
  \bibinfo{pages}{26444--26453} (\bibinfo{year}{2016}).

\bibitem{Kwiat1994}
\bibinfo{author}{Kwiat, P.~G.}, \bibinfo{author}{Steinberg, A.~M.},
  \bibinfo{author}{Chiao, R.~Y.}, \bibinfo{author}{Eberhard, P.~H.} \&
  \bibinfo{author}{Petroff, M.~D.}
\newblock \bibinfo{journal}{\bibinfo{title}{Absolute efficiency and
  time-response measurement of single-photon detectors}}.
\newblock {\emph{\JournalTitle{Applied optics}}} \textbf{\bibinfo{volume}{33}},
  \bibinfo{pages}{1844--1853} (\bibinfo{year}{1994}).

\bibitem{Nomerotski2019}
\bibinfo{author}{Nomerotski, A.}
\newblock \bibinfo{journal}{\bibinfo{title}{Imaging and time stamping of
  photons with nanosecond resolution in timepix based optical cameras}}.
\newblock {\emph{\JournalTitle{Nuclear Instruments and Methods in Physics
  Research Section A: Accelerators, Spectrometers, Detectors and Associated
  Equipment}}} \textbf{\bibinfo{volume}{937}}, \bibinfo{pages}{26--30}
  (\bibinfo{year}{2019}).

\bibitem{ianzano_fast_2020}
\bibinfo{author}{Ianzano, C.} \emph{et~al.}
\newblock \bibinfo{journal}{\bibinfo{title}{Fast camera spatial
  characterization of photonic polarization entanglement}}.
\newblock {\emph{\JournalTitle{Scientific Reports}}}
  \textbf{\bibinfo{volume}{10}}, \bibinfo{pages}{6181},
  \doiprefix\url{10.1038/s41598-020-62020-z} (\bibinfo{year}{2020}).
\newblock \bibinfo{note}{Number: 1 Publisher: Nature Publishing Group}.

\bibitem{Frojdh2015}
\bibinfo{author}{Frojdh, E.} \emph{et~al.}
\newblock \bibinfo{journal}{\bibinfo{title}{Timepix3: first measurements and
  characterization of a hybrid-pixel detector working in event driven mode}}.
\newblock {\emph{\JournalTitle{Journal of Instrumentation}}}
  \textbf{\bibinfo{volume}{10}}, \bibinfo{pages}{C01039}
  (\bibinfo{year}{2015}).

\bibitem{Meduna2019}
\bibinfo{author}{Meduna, L.} \emph{et~al.}
\newblock \bibinfo{journal}{\bibinfo{title}{Real-time timepix3 data clustering,
  visualization and classification with a new clusterer framework}}.
\newblock {\emph{\JournalTitle{arXiv preprint arXiv:1910.13356}}}
  (\bibinfo{year}{2019}).

\bibitem{Kim2020}
\bibinfo{author}{Kim, G.}, \bibinfo{author}{Park, K.}, \bibinfo{author}{Lim,
  K.}, \bibinfo{author}{Kim, J.} \& \bibinfo{author}{Cho, G.}
\newblock \bibinfo{journal}{\bibinfo{title}{Improving spatial resolution by
  predicting the initial position of charge-sharing effect in photon-counting
  detectors}}.
\newblock {\emph{\JournalTitle{Journal of Instrumentation}}}
  \textbf{\bibinfo{volume}{15}}, \bibinfo{pages}{C01034--C01034},
  \doiprefix\url{10.1088/1748-0221/15/01/c01034} (\bibinfo{year}{2020}).

\bibitem{Sajeed2021}
\bibinfo{author}{Sajeed, S.} \& \bibinfo{author}{Jennewein, T.}
\newblock \bibinfo{journal}{\bibinfo{title}{Observing quantum coherence from
  photons scattered in free-space}}.
\newblock {\emph{\JournalTitle{Light: Science \& Applications}}}
  \textbf{\bibinfo{volume}{10}}, \bibinfo{pages}{1--9} (\bibinfo{year}{2021}).

\end{thebibliography}

\end{document}